\documentclass[showpacs,amsmath,amssymb,preprint,aps]{revtex4}
\usepackage[T1]{fontenc}
\usepackage{amstext}
\begin{document}

\title{Electric fields at finite temperature}
\author{A. D. Berm\'{u}dez Manjarres}
\email{ ad.bermudez168@uniandes.edu.co  }
\author{N. G. Kelkar}
\email{nkelkar@uniandes.edu.co}
\author{M. Nowakowski}
\email{mnowakos@uniandes.edu.co}
\affiliation{Departamento de F\'{i}sica\\
Universidad de los Andes\\
Cra. 1E No. 18A-10
Bogot\'{a}, Colombia}
\begin{abstract}
Partial differential equations for the electric potential
at finite temperature, taking into account the thermal Euler-Heisenberg
contribution to the electromagnetic Lagrangian are derived. 
This complete temperature dependence
introduces quantum corrections to several well known equations such as the Thomas-Fermi
and the Poisson-Boltzmann equation. Our unified approach allows at the same time
to derive other similar equations which take into account the effect of the
surrounding heat bath on electric fields. 
We vary our approach by considering a neutral 
plasma as well as the screening caused by electrons only. 
The effects of changing the statistics from Fermi-Dirac to the Tsallis 
statistics and including the presence of a magnetic field are also investigated.
Some useful applications of the above formalism are presented. 
\end{abstract}
\pacs{11.10.Wx,12.20.Ps,03.50.De,26.20.-f}
\maketitle

\section{Introduction}
A class of nonlinear Poisson equations of the form $\mathbf{\mathbf{\nabla}}^2\Phi=F(\Phi,T, \mathbf{r})$ (with $F$ a function)
which take into account the effects (like the temperature dependence $T$) of the surrounding matter on the electric potential $\Phi$
play an important role in many branches of physics. We mention here the Thomas-Fermi equation which finds
its applications in atomic physics \cite{TFatomic}, astrophysics \cite{TFastro} and solid states physics \cite{TFsolids}
and Poisson-Boltzmann equation applied in plasma physics \cite{PBplasma} and solutions \cite{PBsolutions}.  
The derivation of these equations is seemingly unrelated and yet, as shown in this work, they are based on one and the same
principle. Feynman, Metropolis and Teller \cite{Feynman} have have postulated a self-consistent Poisson-like equation of the form 
$\mathbf{\mathbf{\nabla}}^2 \Phi=\int d^3p FD(\Phi, T, \mathbf{p})$,
where $FD$ stands for the Fermi-Dirac distribution, from which the Thomas-Fermi, Poisson-Boltzmann and other similar equations can be derived.
We use this unifying principle to calculate quantum corrections to these nonlinear Poisson equations. These corrections arise when we use the
Quantum Electrodynamics (QED) at finite temperature to calculate the first quantum corrections to the classical electrodynamics known as the Euler-Heisenberg
theory (in our case at finite $T$). The Euler-Heisenberg theory at finite $T$ brings yet another temperature dependence of the electric potential.
To be specific, 
the QED
effective Lagrangian in the presence of a thermal
bath and arbitrary slowly varying electromagnetic field can be written
as
\begin{equation}
\mathcal{L}=\frac{E^{2}-B^{2}}{8\pi}+\mathcal{L}_{EH}^{0}(\mathbf{E},\mathbf{B})+\mathcal{L}_{EH}^{T}(\mathbf{E},\mathbf{B};T),\label{Eff L}
\end{equation}
where $\mathcal{L}_{EH}^{0}(\mathbf{E},\mathbf{B})$ is the zero temperature 
effective Lagrangian of QED \cite{EH,weisskopf,Schwinger} giving rise to new effects like vacuum birefringence \cite{A1,A2,Kruglov}, vacuum dichroism \cite{A3},  
corrections to the Lorentz force \cite{A32}, corrections to the field and energy of point charges \cite{Costa,Kruglov2} among others (see \cite{A4,A5} for comprehensive reviews);  and $\mathcal{L}_{EH}^{T}(\mathbf{E},\mathbf{B};T)$
is the contribution from the thermal bath to the effective Lagrangian \cite{Dittrich -1}. The
Lagrangian (\ref{Eff L}) gives rise to modifications of the Maxwell's equations
that can be used to study electromagnetic phenomena that occur 
beyond the classical electrodynamics \cite{prapaper}.

The initial investigation about the finite temperature effective Lagrangian
was done by Dittrich \cite{Dittrick first FT}. Further developments 
were made in \cite{FT2,FT3,FT4,strong field FT,FT5,two loop,Hight T L,Hight T L2}.
In particular, Ref \cite{two loop} was the first one to show that,
at temperatures below 
$m_{e}$ (mass of the electron), the two loop contributions dominate 
over the one loop term. A review and an expanded bibliography can
be found in \cite{Dittrich -1}.

Among the applications of the finite temperature Lagrangian, we can
find the study of thermally induced photon splitting\cite{PS ,PS 2},
thermally induced pair production \cite{pair1,pair2,pair3}, and the
velocity shift of light in thermalized media \cite{Light 1,light 2,light 3}
(see \cite{Dittrich -1} for more references).

Classical (or semi-classical) methods have been developed for the study
of matter at laboratory conditions or 
plasmas in stars, supernovas, and even the electron-positron
plasma at an early stage of the big-bang \cite{electron-positron}.
We refine these methods by including the effects of 
the QED effective Lagrangian (\ref{Eff L}).  
We do so by implementing the effects of the Euler-Heisenberg theory via the modified Gauss law in the Poisson-like equations.
The set of the latter encompasses known equations 
(like the Thomas-Fermi or Poisson-Boltzmann, now
equipped with quantum corrections) as well as new equations which will be derived in this work.
 
The paper is organized as follows.
In the next section we present the low temperature and high temperature
approximation for the Euler-Heisenberg effective Lagrangian and we will
discuss the way of incorporating the effective Lagrangian into the
equations of classical electrodynamics. 

In section III we calculate the correction to the electrostatic
potential of point-like and extended charged objects when the charge density 
is given. An explicit solution of an electric field at finite temperature
due to the Euler-Heisenberg theory is given. This solution neglects
the fact that the particles surrounding the charge whose 
potential we wish to calculate
can also be in a heat bath. However, the solution is part of a more general
treatment where it appears in the boundary conditions.

In section IV we focus on the temperature dependent charge densities.
We shall write the charge density with two separated terms 
as $\rho=\rho_{c}+\rho_{m}$ where $\rho_{c}$
is the density of the object whose 
effective electrostatic potential we want to compute and $\rho_{m}$ is the charge density
of the surrounding media. 
This results in the Feynman-Metropolis-Teller equation. We discuss several limiting
cases of this master equation treating the degenerate and non-degenerate cases and carefully
distinguishing between the relativistic and non-relativistic situation and the high and low temperature cases. Taking into account
the corrections from the Euler-Heisenberg theory we derive several nonlinear
Poisson-like equations at finite temperature.  

Section V is devoted to the ``relatives'' of the Thomas-Fermi equation, namely equations
derived under a change of assumptions. In the first case we change the Fermi-Dirac
distribution for the Tsallis statistics and the second case considers a Thomas-Fermi
equation in the presence of a magnetic field.

In the section VI we discuss two possible applications, one connected with tunneling
in the presence of a surrounding heat bath and the second one treating an 
electron-positron neutral plasma.

In the last section we draw our conclusions.

\section{Euler Heisenberg Lagrangian for Low and High Temperature}

The full expression of both the zero temperature 
and the thermal Euler-Heisenberg
Lagrangian is very complex. In this work we shall concentrate on some
special cases where the effective Lagrangian can be approximated by 
more manageable expressions. First, we shall deal only with electromagnetic
fields that are weak compared to the so called critical field 
$B_{c}=e^{2}/m_{e}$.
Secondly, all the fields are considered to be slowly varying compared
to the scales of the problem, i.e., the fields obey $\left|\partial_{a}F_{\mu\nu}\right|/\left|F_{\mu\nu}\right|^{2}\ll\left|2F_{\mu\nu}\right|^{1/2}$,
$m_{e}$, $T$ and $\eta^{1/3}$, where $\eta$ is the particle density.
Finally, for the thermal Lagrangian, we will only consider the two
limiting cases of temperature much bigger or much smaller than the
electron mass. 

With the above restrictions the zero temperature Lagrangian can be
written as \cite{EH,Schwinger,weisskopf},  
\begin{equation}
\mathcal{L}_{EH}^{0}=a\left(4\mathcal{F}^{2}+7\mathcal{G}^{2}\right),
\end{equation}
where 
\begin{equation}
a=\frac{e^{4}}{360\pi^{2}m_{e}^{4}},
\end{equation}
and the two relativistic invariants of the electromagnetic fields are
given by
\begin{eqnarray}\label{InvariantG}
\mathcal{F} & = & -\frac{E^{2}-B^{2}}{2},\\\label{InvariantF}
\mathcal{G} & = & \mathbf{E\cdot B}.
\end{eqnarray}

As mentioned in the introduction, for temperatures below the electron
mass $(T\ll m_{e})$, the dominant contribution in the thermal Lagrangian comes
from the two loop term \cite{two loop}. To quadratic order in the
field invariants, the weak field expansion for the two loop Lagrangian
is \cite{two loop}, 

\begin{eqnarray}
\mathcal{L}_{EH}^{T}(T\ll m_{e}) & = & b\left(\mathcal{F}+\mathcal{E}\right)-c\mathcal{F}\left(\mathcal{F}+\mathcal{E}\right)\nonumber \\
 &  & +k\left(2\mathcal{F}^{2}+6\mathcal{F}\mathcal{E}+3\mathcal{E}^{2}-\mathcal{G}^{2}\right).\label{Low L}
\end{eqnarray}
The coefficients appearing in (\ref{Low L}) are 
\begin{eqnarray}
b & = & \frac{44\alpha^{2}\pi^{2}}{2025}\frac{T^{4}}{m_{e}^{4}},\\
c & = & \frac{2^{6}\times37\alpha^{3}\pi^{3}}{3^{4}\times5^{2}\times7}\frac{T^{4}}{m_{e}^{8}},\\
k & = & \frac{2^{13}\alpha^{3}\pi^{5}}{3^{6}\times5\times7^{2}}\frac{T^{6}}{m_{e}^{10}},
\end{eqnarray}
 and $\mathcal{E}$ is a term involving the relative velocity of
the thermal bath. We will work only in the reference frame where the
bath is at rest, and in that special frame we have $\mathcal{E}=E^{2}$. 

In the high temperature limit $(T\gg m_{e})$ the one loop correction
is the dominating term and the thermal correction takes the form \cite{Hight T L,Hight T L2}
\begin{equation}
\mathcal{L}_{EH}^{T}(T\gg m_{e})=-\frac{2\alpha}{3\pi}\mathcal{F}\ln\left(\frac{T}{m_{e}}\right)+\frac{\alpha}{6\pi}\mathcal{E}-\mathcal{L}_{EH}^{0}.\label{High L}
\end{equation}

It can be seen from the above equation 
(\ref{High L}) that, for temperatures above the
electron mass, the thermal bath cancels the vacuum polarization effects
from the zero temperature Euler-Heisenberg Lagrangian.

In this paper, the interest in the effective Lagrangian comes from
the fact that it can be related to a modification of the Maxwell's
equations at the purely classical level. 

Faraday's and magnetic Gauss's laws remain unchanged 
\begin{eqnarray}
\mathbf{\mathbf{\nabla}}\cdot\mathbf{B} & = & 0,\label{mgauss}\\
\mathbf{\mathbf{\nabla}}\times\mathbf{E} & = & -\frac{\partial\mathbf{B}}{\partial t}.\label{Faradays}
\end{eqnarray}

We see from (\ref{mgauss}) and (\ref{Faradays}) that electromagnetic
potentials are still defined as in classical electrodynamics i.e,
$\mathbf{E}=-\mathbf{\nabla}\phi-\frac{\partial\mathbf{A}}{\partial t}$ and
$\mathbf{B}=\mathbf{\nabla}\times\mathbf{A}$.

The electric Gauss's and Ampere-Maxwell's law are modified by the
use of the effective Lagrangian. They now resembles the the form of
the Maxwell's equation in matter \cite{Landau}, namely, 
\begin{eqnarray}
\mathbf{\nabla}\cdot\mathbf{D} & = & 4\pi\rho,\label{Egauss}\\
\mathbf{\nabla}\times\mathbf{H} & = & \frac{\partial\mathbf{D}}{\partial t},\label{amperes}
\end{eqnarray}
where $\rho$ is the charge density and the auxiliary fields $\mathbf{D}$
and $\mathbf{H}$ are given by,  
\begin{eqnarray}
\mathbf{D} & = & \mathbf{E}+4\pi\frac{\partial\mathcal{L}_{EH}}{\partial\mathbf{E}},\\
\mathbf{H} & = & \mathbf{B}-4\pi\frac{\partial\mathcal{L}_{EH}}{\partial\mathbf{B}}.
\end{eqnarray}

In classical electrodynamics the Gauss law $\mathbf{\nabla}\cdot\mathbf{E}=\rho$
or Laplace equation $-\mathbf{\nabla}^{2}\phi=\rho$ is solved in order to
find the field created by a given charge density distribution. Here
we shall tackle the problem of finding the effective electric field of
a spherical charge distribution produced by the 
modified Gauss Law (\ref{Egauss}).

\section{Temperature Independent charge density}
To begin with, we shall derive the fields in the limit of low and high 
temperatures for the case of temperature independent charge densities.  
\subsection{Low temperature}

For pure electric field ($\mathbf{B}=0$) and in the plasma rest frame,
the effective Lagrangian takes the form
\begin{eqnarray}
\mathcal{L}_{Maxwell}+\mathcal{L}_{EH}^{0}+\mathcal{L}_{EH}^{T} & = & (\frac{1}{8\pi}+\frac{b}{2})E^{2}+(a+\frac{k}{2}-\frac{c}{4})E^{4}.
\end{eqnarray}
With this Lagrangian the Gauss law reads 
\begin{equation}
\mathbf{\nabla}\cdot(A(T)E^{2}\mathbf{E}+B(T)\mathbf{E})=4\pi
\rho(\mathbf{r})\label{lowTE}
\end{equation}
with 
\begin{eqnarray}
A(T) & = & 16\pi\left(a+\frac{k}{2}-\frac{c}{4}\right),\label{A(T)}\\
B(T) & = & 1+4\pi b.\label{B(T)}
\end{eqnarray}
The rest of this sections follows closely the works \cite{Costa,Kruglov2}. In general, The charge distribution can be written as
\begin{equation}
4\pi\rho(\mathbf{r})=\mathbf{\nabla}\cdot\mathbf{E}_{c}\label{Ec}
\end{equation}
where $\mathbf{E}_{c}$ is the field that would be produced by $\rho$
in Maxwell's theory. 

The field $\mathbf{E}$ given by (\ref{lowTE}) has to approach $\mathbf{E}_{c}$
in the limit that the Euler-Heisenberg coefficients vanish. We can
then write from (\ref{lowTE}) and (\ref{Ec}) the following algebraic
equation
\begin{equation}
A(T)E^{3}+B(T)E=E_{c}.\label{algebriaclowt}
\end{equation}
This equation is a cubic equation which has only
one real solution and is given by Cardano's formula, 
\begin{equation}
E=\sqrt[3]{-\frac{E_{c}}{A(T)}+\sqrt{\frac{E_{c}^{2}}{A^{2}(T)}+\frac{B^{3}(T)}{A^{3}(T)}}}+\sqrt[3]{-\frac{E_{c}}{A(T)}-\sqrt{\frac{E_{c}^{2}}{A^{2}(T)}+\frac{B^{3}(T)}{A^{3}(T)}}}.
\label{cardanos}
\end{equation}

Let's note that at large distances, the behaviour of $E$ is
\begin{equation}
E\sim\frac{e}{B(T)r^{2}}\label{long E}
\end{equation}
while for short distances it is
\begin{equation}
E\sim\left[\frac{e}{A(T)r^{2}}.\right]^{1/3}.\label{short E}
\end{equation}

In particular, we will later need the form of the potential at short
distances. The behaviour of the potential for small $r$ follows from  
(\ref{short E}) to be 
\begin{equation}
\phi\sim-\frac{1}{3}\left[\frac{e}{A(T)}\right]^{1/3}r^{1/3}+\phi(0)\label{phi0}
\end{equation}
where $\phi(0)$ is a positive constant given by
\begin{equation}
\phi(0)=-\int_{0}^{\infty}Edr.
\end{equation}

Equation (\ref{cardanos}) would be a complete result if we could neglect screening
effects and neglect the temperature dependence of the surrounding matter.
Nevertheless the result (\ref{cardanos}) is important for later, 
more exhaustive considerations
when we will take into account the temperature dependence of the matter. 
Then the boundary condition of the resulting partial differential equations 
will be formulated with the help of (\ref{cardanos}), i.e., at small distances
the potential should follow (\ref{phi0}).

\subsection{High Temperature}

The high temperature case is mathematically easier. 
\begin{eqnarray*}
\mathbf{\nabla}\cdot\mathbf{E}_{c}=\rho(\mathbf{r}) & = & \mathbf{\nabla}\cdot\mathbf{D}=\mathbf{\nabla}\cdot(\mathbf{E}+4\pi\mathbf{P})\\
 & = & \left[\frac{8\alpha}{3}\ln\left(\frac{T}{m_{e}}\right)+\frac{4\alpha}{3}+1\right]\mathbf{E}
\end{eqnarray*}
The solution for $\mathbf{E}$ is trivial if $\mathbf{E}_{c}$ is
known and is given by 
\[
\mathbf{E}=\frac{1}{\frac{8\alpha}{3}\ln\left(\frac{T}{m_{e}}\right)+\frac{4\alpha}{3}+1}\mathbf{E}_{c}
\]

\section{Temperature Dependent charge density}

We will work out the effective electric field of a point charge that
is submerged in a neutral plasma consisting of electrons and positive
charges which can consist of protons or positrons.

As is customary, statistical methods are needed when the charge density
depends on temperature. The density of fermions is governed by 
the Fermi-Dirac distribution and is given by
\begin{equation}
\eta(\mathbf{r},T)=4(2\pi)^{4}\int_{0}^{\infty}\frac{\left(p^{2}/\hbar^{3}\right)dp}{e^{\beta\left(K+q\phi+\mu\right)}+1} \,,\label{eq:density}
\end{equation}
where $\mu$ stands for the chemical potential, $\hbar$ denotes
the Planck's constant, we use $c = 1$ and $K$ is the kinetic energy
\begin{equation}
K=\begin{cases}
p^{2}/2m & non-relativistic,\\
\sqrt{p^{2}+m^{2}} & relativistic,\\
p & ultra-relativistic.
\end{cases}
\end{equation}
The electron charge density is then given by $-e\eta_{e}(\mathbf{r},T)$
and an analogous analysis holds for the proton or positron charge
density.

The equation for the effective field 
at $\mathbf{r}$ created by a charge distribution 
is the modified Poisson's equation
\begin{equation}
\mathbf{\nabla}\cdot\left(\mathbf{\mathbf{E}+4\pi\frac{\partial\mathcal{L}_{EH}}{\partial\mathbf{E}}}\right)=\rho(\mathbf{r},T)_{total},\label{D density}
\end{equation}
where the term $\rho(\mathbf{r},T)_{total}$ means the density taking
into account all the electric charges and it depends on the system
considered. For example, consider the densities given by
\begin{eqnarray}
\rho(\mathbf{r},T)_{total} & = & -e\eta_{e}(\mathbf{r},T),\\
\rho(\mathbf{r},T)_{total} & = & \rho(\mathbf{r})-e\eta_{e}(\mathbf{r},T),\\
\rho(\mathbf{r},T)_{total} & = & \rho(\mathbf{r})+4\pi e\eta_{p}(\mathbf{r},T)-4\pi e\eta_{e}(\mathbf{r},T).
\end{eqnarray}

The above densities stand respectively for (a) a cloud of electron, (b)
a given particle with charge density $\rho(\mathbf{r}$) that is surrounded
by a cloud of electrons and (c) a charge $\rho(\mathbf{r}$)
surrounded by a cloud of electrons and protons. We always consider
the charges to have spherical symmetry. Neglecting the Euler-Heisenberg contribution 
in (\ref{D density}) and introducing an electric potential $\phi$, 
the resulting equation
\begin{equation}\label{FMT}
\mathbf{\nabla}^2 \phi = \rho(\mathbf{r},T)_{total}\, ,
\end{equation}
is referred to as the Feynman-Metropolis-Teller (FMT) equation.

Using equation (\ref{D density}) may prove to be too difficult as
stated. It is convenient to look for special cases 
where the density (\ref{eq:density})
can be reduced to a simpler expression. The way to simplify (\ref{eq:density})
depends on the relation between the temperature and the chemical potential
in a given situation.

At this point it is important to make some clarifications about what
we mean by the temperature regime. The problem at hand has two natural
scales of temperatures and they enter in different ways in the modified
Poisson's equation (\ref{D density}). First, we can compare $T$ with
$m_{e}$, and as we have seen this affects the form of the effective
Lagrangian e.g, for $T\gg m_{e}$ we use (\ref{High L}) in the left
hand side of (\ref{D density}) and for $T\ll m_{e}$ we use the Lagrangian
(\ref{Low L}) instead. On the other hand, for the right hand side
of (\ref{eq:density}) the temperature has to be compared with 
the chemical potential $\mu$ in order to know what simplification
can be done to the charge density. We have a degenerate system when
$T\ll\mu$ and a dilute system when $T\gg\mu$, both with different
approximate expansions (the form of the kinetic energy also affects
the approximation of (\ref{eq:density})).

In the appendix we summarize different ways to approximate equation
(\ref{eq:density})

\subsection{Low temperature}

We deal first with the systems whose temperatures are below $m_{e}$
and are composed of electrons and protons at equilibrium. The equation
of interest is 
\begin{eqnarray}
\mathbf{\nabla}\cdot(A(T)E^{2}\mathbf{E}+B(T)\mathbf{E}) & = & 
4\pi\rho(\mathbf{r},T)_{total},\label{lowtphi}
\end{eqnarray}

Since the charge density is related to the electric potential, it is better
to work with the potential instead of the electric field. To accomplish
the above we first mention that the usual relation between the electric
field and the potential is maintained i.e, $\mathbf{E}=-\mathbf{\nabla}\phi$.
Then, it is just a matter of replacing the potential for electric field
into (\ref{lowtphi}). Taking into account the spherical symmetry
of the problem, the differential equation for the potential of a charge
distribution can be written as
\begin{eqnarray}
-\left(\mathbf{\nabla}^{2}+\mathbf{\mathbf{\mathfrak{D}}}_{EH}^{2}\right)\phi & =4\pi\rho(\mathbf{r},T)_{total} & ,
\end{eqnarray}
where the non-linear differential operator $\mathbf{\mathfrak{D}}_{EH}^{2}$
stands for
\begin{equation}
\mathbf{\mathbf{\mathfrak{D}}}_{EH}^{2}\bullet=A(T)\frac{1}{r^{2}}\left(\frac{d\bullet}{dr}\right)^{2}\frac{d}{dr}\left(r^{2}\frac{d\bullet}{dr}\right)+A(T)\left(\frac{d\bullet}{dr}\right)^{2}\frac{d^{2}\bullet}{dr^{2}}+4\pi b\mathbf{\nabla}^{2}\bullet,\label{shorthand}
\end{equation}
with $A(T)$ and $B(T)$ given by (\ref{A(T)}) and (\ref{B(T)}).
Let's note that $-\mathbf{\mathfrak{D}}^{2}\rightarrow0$ when $\alpha\rightarrow0$.

\subsubsection{Degenerate matter}

Before we embark on the derivation of the Thomas-Fermi like differential equations
we note that for astrophysical applications  a rough distinction between 
degenerate and non-degenerate matter can be made according to the 
mass density $\rho$ \cite{Hansen}. For non-degenerate matter we should have
\begin{equation} \label{Hansen1}
\rho \le 8.49 \times 10^{-17}\left(\frac{T^3}{1K}\right) \frac{\rm g}{\rm cm^3}
\end{equation}
whereas the degenerate matter should satisfy
\begin{equation} \label{Hansen2}
\rho \ge 4.2 \times 10^{-10}\left(\frac{T^3}{1K}\right) 
\frac{\rm g}{\rm cm^3}\, .
\end{equation}

We now present the differential equations for the potentials in the
case where the temperature is low compared to the chemical potential. 

For the degenerate case we can use the particle density 
(\ref{non-rel degenerate}) (given in the appendix) 
to write the non-relativistic equation as
\begin{eqnarray}
 & -\left(\mathbf{\nabla}^{2}+\mathbf{\mathfrak{D}}_{EH}^{2}\right)\phi=\rho(\mathbf{r})\nonumber \\
 & +\frac{4(2\pi)^{5}e(2m_{p}T){}^{3/2}}{\hbar^{3}}\left(\frac{2}{3}\left(\frac{\mu_{p}-e\phi}{T}\right)^{3/2}+\frac{\pi^{2}}{12}\left(\frac{\mu_{p}-e\phi}{T}\right)^{-1/2}\right)\nonumber \\
 & -\frac{4(2\pi)^{5}e(2m_{e}T){}^{3/2}}{\hbar^{3}}\left(\frac{2}{3}\left(\frac{\mu_{e}+e\phi}{T}\right)^{3/2}+\frac{\pi^{2}}{12}\left(\frac{\mu_{e}+e\phi}{T}\right)^{-1/2}\right).\label{TFNR}
\end{eqnarray}

Using (\ref{ult-rel degenerate}) (from the appendix) 
the ultra-relativistic equation is
\begin{eqnarray}
 & -\left(\mathbf{\nabla}^{2}+\mathbf{\mathfrak{D}}_{EH}^{2}\right)\phi=\rho(\mathbf{r})\nonumber \\
 & +\frac{8(2\pi)^{5}e}{\hbar^{3}}\left(\mu_{p}-e\phi\right)^{3}\left(1+\frac{3}{4}\pi^{2}\left(\frac{T}{\mu_{p}-e\phi}\right)^{2}\right)\nonumber \\
 & -\frac{8(2\pi)^{5}e}{\hbar^{3}}\left(e\phi+\mu_{e}\right)^{3}\left(1+\frac{3}{4}\pi^{2}\left(\frac{T}{e\phi+\mu_{e}}\right)^{2}\right).\label{TMUR}
\end{eqnarray}
Equations (\ref{TFNR}) and (\ref{TMUR}) are the low $T$ Euler-Heisenberg
generalization to the non-relativistic and ultra-relativistic Thomas-Fermi
equations with the first thermal correction and a contribution from
positive charges. As already mentioned in section III we can 
drop the source $\rho$ from these equations (and the other equations
derived below) by incorporating it into the boundary conditions.
This is to say, we demand that at small distances the potential behaves as
(\ref{phi0}) which amounts to saying that the effect of matter is negligible.
In the standard Thomas-Fermi-like equations for spherical charge 
distributions and without the Euler-Heisenberg
corrections this is equivalent to demanding that the Coulomb law 
be valid at short distances.

For the sake of comparison we now present the standard Thomas-Fermi
equations for the non-relativistic and the ultra-relativistic cases.
Considering a point charge $Z_{1}e$ surrounded by a cloud of electrons, the
non-relativistic Thomas-Fermi equation with the first thermal term 
is \cite{Bethe}
\begin{equation}
\mathbf{\nabla}^{2}\phi=\frac{8e(2\pi)^{5}(2m_{e}){}^{3/2}}{3\hbar^{3}}\left(\mu_{e}+e\phi\right)^{3/2}\left(1+\frac{\pi^{2}}{8}\left(\frac{\mu_{e}+e\phi}{T}\right)^{-2}\right).\label{OTF}
\end{equation}
The ultra-relativistic Thomas-Fermi equation reads
\begin{equation}
\mathbf{\nabla}^{2}\phi=\frac{4(2\pi)^{4}e}{\hbar^{3}}\left(\mu+e\phi\right)^{3}\left(1+\frac{3}{4}\pi^{2}\left(\frac{T}{\mu+e\phi}\right)^{2}\right).\label{OUTF}
\end{equation}

It is possible to rewrite the Thomas-Fermi equation (\ref{OTF}) using
the following change of variables 
\begin{eqnarray}
\mu_{e}+e\phi & = & \frac{Z_{1}e^{2}}{r}\psi(s),\label{CV1}\\
r & = & Cs,\label{CV2}\\
C & = & \frac{1}{2}\left(\frac{3}{8\left(2\pi\right)^{5}}\right)^{2/3}Z_{1}^{-1/3}a_{0},\label{CV3}
\end{eqnarray}
where $a_{0}=\hbar^{2}/e^{2}m_{e}$ is the Bohr radius and
$s$ is a dimensionless quantity. 

After using (\ref{CV1}), (\ref{CV2}) and (\ref{CV3}), the Thomas-Fermi
equation (\ref{OTF}) becomes 
\begin{equation}
\frac{d^{2}\psi(s)}{ds^{2}}=\frac{\psi(s)^{3/2}}{\sqrt{s}}\left(1+\frac{\gamma T^{2}s^{2}}{\psi^2(s)}\right),\label{TF}
\end{equation}
where $\gamma=\frac{\pi^{2}}{8}\left(\frac{3}{8\left(2\pi\right)^{5}}\right)^{2/3}Z_{1}^{-1/3}\frac{a_{0}^{2}}{e^{4}}$. 

Another customary way to rewrite equation (\ref{OTF}) is by using the
change of variables \cite{Feynman},   
\begin{eqnarray}
\frac{\mu_{e}+e\phi}{T} & = & \frac{\psi}{s},\\
C' & = & \sqrt{\frac{\hbar^{3}}{8e^{2}(2\pi)^{5}(2m_{e}T){}^{1/2}}}.
\end{eqnarray}
In this case $\psi$ obeys
\begin{equation}
\frac{d^{2}\psi(s)}{ds^{2}}=\frac{\psi(s)^{3/2}}{\sqrt{s}}\left(1+\frac{\pi^{2}s^{2}}{8\psi(s)^{2}}\right).
\end{equation}

With the definitions (\ref{CV1}), (\ref{CV2}), and (\ref{CV3}),
we can write the low T Euler-Heisenberg generalization to the Thomas-Fermi
equation (\ref{TF}) in the form 
\begin{equation}
\frac{d^{2}\psi(s)}{ds^{2}}+\frac{1}{C^{4}}\mathbf{\mathfrak{D}}'{}_{EH}^{2}\psi(s)=\frac{\psi(s)^{3/2}}{\sqrt{s}}\left(1+\frac{\gamma T^{2}s^{2}}{\psi(s)^{2}}\right),\label{TF-1}
\end{equation}
where $\mathbf{\mathfrak{D}}'{}_{EH}^{2}$ in terms of $\psi$ and $s$ reads
\begin{eqnarray}
\mathbf{\mathfrak{D}}'{}_{EH}^{2}\bullet & = & A(T)\left[\frac{1}{s}\frac{d\bullet}{ds}-\frac{\bullet}{s^{2}}\right]^{2}\nonumber \\
 &  & \times\left[2\frac{d^{2}\bullet}{ds^{2}}-2\frac{1}{s}\frac{d\bullet}{ds}+2\frac{\bullet}{s^{2}}\right].
\end{eqnarray}

The equation (\ref{TFNR}) that considers both a cloud of negative
and positive charges can also be rewritten using the change of variables 
(\ref{CV1}) and (\ref{CV2}), with the result 
\begin{eqnarray}
\frac{d^{2}\psi(s)}{ds^{2}}&+&\frac{1}{C^{4}}\mathbf{\mathfrak{D}}'{}_{EH}^{2}\psi(s) =  -s\rho(Cs)\nonumber \\
 &  +&\frac{\psi(s)^{3/2}}{\sqrt{s}}\left(1+\frac{\gamma T^{2}s^{2}}{\psi(s)^{2}}\right)\nonumber \\
 &   +&\frac{8e(2\pi)^{5}(2m_{e}){}^{3/2}C^{2}}{3\hbar^{3}}s\nonumber \\
 & \times &\left(\frac{e^{2}\psi(s)}{Cs}+\mu_{+}-\mu_{e}\right)^{3/2}\left(1+\frac{\pi^{2}T^{2}}{8}\left(\frac{e^{2}\psi(s)}{Cs}+\mu_{+}-\mu_{e}\right)^{-2}\right).\label{TFNR-1}
\end{eqnarray}
We can see that the whole effect of the low $T$ Euler-Heisenberg
is contained in the term $\frac{1}{C^{4}}\mathbf{\mathfrak{D}}'{}_{EH}^{2}\psi$.

A similar treatment can be given for the ultra-relativistic equations.
The change of variables is the same as (\ref{CV1}) and (\ref{CV2})
but with $C=\sqrt{\frac{\hbar^{3}}{4(2\pi)^{4}e}}$ (though in this
case $s$ is not dimensionless). With this change of variables equation
(\ref{OUTF}) becomes
\begin{equation}
\frac{d^{2}\psi(s)}{ds^{2}}=\frac{\psi(s)^{3}}{s^{2}}\left(1+\frac{3\pi^{2}T^{2}s^{2}}{4\psi(s)^{2}}\right).\label{UTF1}
\end{equation}

The Euler-Heisenberg generalization to (\ref{UTF1}) now reads 
\begin{equation}
\frac{d^{2}\psi(s)}{ds^{2}}+\frac{1}{C^{4}}\mathbf{\mathfrak{D}}'{}_{EH}^{2}\psi(s)=\frac{\psi(s)^{3}}{s^{2}}\left(1+\frac{3\pi^{2}T^{2}s^{2}}{4\psi(s)^{2}}\right).\label{UTF2}
\end{equation}

With the inclusion of the test charge and a cloud of positive charges,
the equation (\ref{TMUR}) becomes
\begin{eqnarray}
 & \frac{d^{2}\psi(s)}{ds^{2}}+\frac{1}{C^{4}}\mathbf{\mathfrak{D}}'{}_{EH}^{2}\psi(s)=-s\rho(Cs)+\frac{\psi(s)^{3}}{s^{2}}\left(1+\frac{3\pi^{2}}{4}\frac{s^{2}}{\psi(s)^{2}}\right)-\nonumber \\
 & \frac{1}{s^{2}}\left(\frac{e^{2}\psi(s)}{s}+\mu_{+}-\mu_{e}\right)^{3}\left(1+\frac{3}{4}\pi^{2}T^{2}\left(\frac{e^{2}\psi(s)}{s}+\mu_{+}-\mu_{e}\right)^{-2}\right).
\end{eqnarray}

In the standard Thomas-Fermi theory, the function $\psi(r)$ has 
to obey the following boundary conditions
\begin{eqnarray}
\psi(0) & = & 1,\label{cond 1}\\
\psi(\infty) & = & 0,\label{cond 2}
\end{eqnarray}
where we have ignored the size of the charged object. Condition (\ref{cond 1})
ensures that we recover Coulomb electrostatic energy at short distances.
Condition (\ref{cond 2}) ensures the right behaviour at large distances. 
However, for the Euler-Heisenberg generalization of
Thomas-Fermi equations, the potential has to reduce to its Euler-Heisenberg
form (\ref{phi0}) at short distances. Therefore, for small $r$, $\psi$
has to behave like
\begin{equation}
\psi(s)\sim-\frac{1}{3e}\left[\frac{e}{A(T)}\right]^{1/3}s^{7/3}+\phi(0)\frac{s^{2}}{e}.\label{cond 3}
\end{equation}

\subsubsection{Dilute matter}

For dilute matter, the Maxwell's-Boltzmann distribution 
(\ref{classical limit}) given in the appendix, 
can be used to write the density of particles: 
\begin{eqnarray}
n_{e} & = & n_{e0}e^{\frac{e\phi}{T}},\\
n_{p} & = & n_{p0}e^{-\frac{e\phi}{T}},
\end{eqnarray}
where $n_{e0}$ and $n_{i0}$ are the concentrations of electrons and
protons respectively. 

Assuming the concentration of negative and positive charges to be equal
to $n_{0}$, we can write the equation
\begin{eqnarray}
-\left(\mathbf{\mathbf{\mathbf{\nabla}}}^{2}+\mathbf{\mathfrak{D}}_{EH}^{2}\right)\phi & = & \rho(\mathbf{r})-en_{0}e^{\frac{e\phi}{T}}+en_{0}e^{-\frac{e\phi}{T}}.\label{PB}
\end{eqnarray}
Equation (\ref{PB}) is a generalization of the Poisson-Boltzmann
equation.

For regions where the perturbed potential obeys $\phi\ll T$, the
exponentials in (\ref{PB}) can be expanded and keeping 
only the first term, we get, 
\begin{eqnarray}
-\left(\mathbf{\nabla}^{2}+\mathbf{\mathfrak{D}}_{EH}^{2}\right)\phi+\kappa^{2}\phi & = & \rho(\mathbf{r})\label{LPB}
\end{eqnarray}
with, $\kappa$, the Debye parameter, given by
\begin{equation}
\kappa^{2}=\frac{2e^{2}n_{0}}{T}.
\end{equation}

Equation (\ref{LPB}) is a generalization of the linearized version
of the Poisson-Boltzmann or Debye-H\"uckel equation \cite{plasma DH}.
Without considering the Euler-Heisenberg term, the Debye-H\"uckel equation
is
\begin{equation}
-\mathbf{\nabla}^{2}\phi+\kappa^{2}\phi=\rho(\mathbf{r}).\label{Cdebye-huckel}
\end{equation}

For for point charges $\rho(\mathbf{r})=q\delta(\mathbf{r})$, equation
(\ref{Cdebye-huckel}) has the solution
\begin{equation}
\phi=\frac{q}{r}e^{-\kappa r}.
\end{equation}
The screening effect is evident.

\subsection{High Temperature}

\subsubsection{Dilute matter}

In the high temperature regime the particles move ultra-relativistically
with kinetic energy $E=p$ and for the non-degenerate case, charge
densities are given by the distribution (\ref{classical limit}). We
assume the electric interaction between charges to be small as compared
to the temperature so that we can write the differential equation for the
screened potential for a point-charge as
\begin{eqnarray}
\left[\frac{8\alpha}{3}\ln\left(\frac{T}{m_{e}}\right)+\frac{4\alpha}{3}+1\right]\mathbf{\nabla}^{2}\phi & = & q\delta(\mathbf{r})+n_{0}e^{\frac{-e\phi}{T}}-n_{0}e^{\frac{e\phi}{T}}\\
 & \approx & q\delta(\mathbf{r})-\kappa^{2}\phi.\label{URLPBE}
\end{eqnarray}

Equation (\ref{URLPBE}) can easily be rewritten as
\begin{equation}
-\mathbf{\nabla}^{2}\phi+\kappa_{EH}^{2}\phi=q_{EH}(T)\delta(\mathbf{r}),\label{modified poisson-b}
\end{equation}
where we have defined
\begin{eqnarray}
\kappa_{EH}^{2} & = & \frac{\kappa^{2}}{\frac{8\alpha}{3}\ln\left(\frac{T}{m_{e}}\right)+\frac{4\alpha}{3}+1},\label{kT}\\
q_{EH}(T) & = & \frac{q}{\frac{8\alpha}{3}\ln\left(\frac{T}{m_{e}}\right)+\frac{4\alpha}{3}+1}.\label{e(T)}
\end{eqnarray}

We see that for dilute gases at temperatures above the electron mass,
the effects of the Euler-Heisenberg Lagrangian are the renormalization
of the Debye parameter (\ref{kT}) and the electric charge (\ref{e(T)}).

The solution of (\ref{modified poisson-b}) is given by \cite{plasma DH}
\begin{equation}
\phi=\frac{q_{EH}(T)}{r}e^{-\kappa_{EH}r}\, .
\end{equation}
This represents an analytical solution of a Poisson-Boltzmann problem
with Euler-Heisenberg corrections.

\section{Related equations}
Having discussed the Thomas-Fermi and other 
equations at different temperatures and 
densities, we now consider the variants for different statistics and the 
equations obtained in the presence of magnetic fields. 
\subsection{Tsallis Statistics}

Tsallis statistics have been with us for about 30 years now \cite{Tsallis 1}.
It has been applied to physical situations like Euler turbulence 
\cite{tsallis euler turbulence}, gravitating systems \cite{tsallis gravitation},
ferrofluid-like systems \cite{tsallis ferro} and neutron stars \cite{tsallis n stars},
among others. Recently, it has been suggested that the Tsallis statistics
could eventually explain the Lithium anomaly of early nucleosynthesis
\cite{bertulani}.

In the Tsallis statistics for Fermi particles the occupation number
is given by
\begin{equation}
\eta=4(2\pi)^{4}\int_{0}^{\infty}\frac{\left(p^{2}/\hbar^{3}\right)dp}{e_{q}(\beta\left(K+q\phi-\mu\right)+1}\label{n tsallis}
\end{equation}
where $q$ is a real number and
\begin{equation}
e_{q}(x)=\left[1+(1-q)x\right]^{\frac{1}{1-q}}\label{G exp}
\end{equation}
is a generalization of the standard exponential function, which is
recovered in the limit $q\rightarrow1$. 

Density (\ref{n tsallis}) has been used in literature to form a
non-extensive generalization of the Thomas-Fermi equations; in the
nonrelatistic case by \cite{Tsallis TM-1}, and in the relativistic
case by \cite{Tsallis RTF-1}.

The relativistic Poisson equation reads
\begin{eqnarray}
\mathbf{\nabla}^{2}\phi & = & \frac{em_{e}^{3}}{3\pi^{2}\hbar^{2}}\left[\frac{\left(\mu+m_{e}+\phi\right)}{m_{e}^{2}}-1\right]^{3/2}\times\left\{ 1+\frac{3TI_{1}^{(q)}}{m_{e}}\left[\frac{\left(\mu+m_{e}+\phi\right)}{m_{e}^{2}}-1\right]^{-1}\right.\nonumber \\
 &  & \left.+\frac{3TI_{2}^{(q)}}{m_{e}}\left[\frac{\left(\mu+m_{e}+\phi\right)}{m_{e}^{2}}-1\right]^{-2}+\ldots\right\} ,\label{tsallis Rpoisson}
\end{eqnarray}
where the q-generalized Fermi-Dirac integral $I_{k}^{(q)}$ is defined
by
\begin{equation}
I_{k}^{(q)}=q\int_{-\infty}^{\infty}\frac{z^{n}\left[1+(q-1)z\right]^{1/(q-1)}dz}{\left\{ 1+\left[1+(q-1)z\right]^{q/(q-1)}\right\} ^{2}}.\label{q FD integral}
\end{equation}
Numerical evaluation of (\ref{q FD integral}) for different q can
be found in \cite{Tsallis Numerics,Tsallis TM-1}.

With the following change of variables
\begin{eqnarray}
\mu_{e}+e\phi & = & \frac{Z_{1}e^{2}}{r}\psi(s),\label{CV1-1}\\
r & = & Cs,\label{CV2-1}\\
C & = & \left(\frac{9\pi^{2}}{128}\right)^{1/3}Z_{1}^{-1/3}a_{0},\label{CV3-1}
\end{eqnarray}
equation (\ref{tsallis Rpoisson}) transform into the non-extensive
relativistic generalization of the Thomas-Fermi equation
\begin{eqnarray}
\frac{d^{2}\psi}{ds^{2}} & = & \frac{\psi{}^{3/2}}{\sqrt{s}}\left(1+\frac{\gamma T^{2}s^{2}}{\psi}\right)\nonumber \\
 &  & \times\left\{ 1+\chi_{1}\frac{Ts}{\psi}\left[1+\gamma\frac{s}{\psi}\right]^{-1}+\chi_{2}\frac{T^{2}s^{2}}{\psi^{2}}\left[1+\gamma\frac{s}{\psi}\right]^{-2}+\ldots\right\} \label{R tsallis G}
\end{eqnarray}
where
\begin{eqnarray}
\gamma=\left[\frac{4Z_{1}^{2}}{3\pi}\right]^{2/3}\frac{e^{4}}{\hbar^{2}}, & \chi_{1}=\frac{3C}{2e^{2}Z_{1}}I_{1}^{(q)}, & \chi_{2}=\frac{3C^{2}}{8e^{4}Z_{1}^{2}}I_{2}^{(q)}.
\end{eqnarray}

In the limit where the relativistic contribution is neglected ($\gamma\rightarrow0$),
equation (\ref{R tsallis G}) reduces to the following non-relativistic
expression (originally obtained in \cite{Tsallis TM-1})
\begin{equation}
\frac{d^{2}\psi}{ds^{2}}=\frac{\psi{}^{3/2}}{\sqrt{s}}\left[1+\chi_{1}\frac{Ts}{\psi}+\chi_{2}\frac{T^{2}s^{2}}{\psi^{2}}\right].\label{NR tsallis G}
\end{equation}

\subsection{Thomas-Fermi equations in presence of magnetic fields}

In the context of nuclear astrophysics, there are cases where 
the process of interest occurs in presence of magnetic fields. When
the magnetic field is intense enough, the quantum nature of the motion
of the charged particle can not be ignored. 

The first investigation of the modification of the Thomas-Fermi equation
due to a magnetic field was done in \cite{MTF 1}. Further developments 
were done in \cite{MTF2,MTF3,MTF4}.
We follow the procedure of \cite{MTF5}, where the discretization
of the transverse motion into Landau levels is taken into account.
The motion of electrons perpendicular to the magnetic field is quantized
into the discrete Landau Levels $\nu B$, with $\nu=0,1,2,...$.
The degeneracy of the levels, per unit area, is $\frac{B}{2\pi}$
for $\nu=0$, but, due to the electron spin, the degeneracy is twice
as high for the higher $\nu$. Along the direction of the field the
motion is not quantized, and the degeneracy of states is $D(\varepsilon)=\varepsilon^{-1/2}/(2^{1/2}\pi)$,
where $\varepsilon$ is the energy of the translational motion.

Taking the above into consideration, it follows that the density of
electrons at temperature $T$ and electrical potential $-e\phi$ is
given by
\begin{eqnarray}
\eta & = & \frac{B}{2\pi}\frac{1}{2^{1/2}\pi}\int_{0}^{\infty}\left[\int_{0}^{\infty}\frac{\varepsilon^{-1/2}}{e^{(\varepsilon-\mu-e\phi)/T}+1}d\varepsilon+2\sum_{v=1}^{\infty}\int_{0}^{\infty}\frac{\varepsilon^{-1/2}}{e^{(\varepsilon+\nu B-\mu-e\phi)/T}+1}d\varepsilon\right]\nonumber \\
 & = & \frac{BT^{1/2}}{2^{3/2}\pi^{2}}\left[I_{-1/2}\left(\frac{\mu+e\phi}{T}\right)+2\sum_{v=1}^{\infty}I_{-1/2}\left(\frac{\mu+e\phi-\nu B}{T}\right)\right]\label{Mdensity}
\end{eqnarray}
where the Fermi-Dirac integral for $k>-1$ is defined by
\begin{equation}
I_{k}(x)=\int_{0}^{\infty}\frac{y^{k}}{e^{y-x}+1}dy.
\end{equation}

Combining (\ref{Mdensity}) with the Poisson's equation yields
\begin{equation}
\mathbf{\nabla}^{2}\phi=4\frac{BT^{1/2}}{2^{3/2}\pi}\left[I_{-1/2}\left(\frac{\mu+e\phi}{T}\right)+2\sum_{v=1}^{\infty}I_{-1/2}\left(\frac{\mu+e\phi-\nu B}{T}\right)\right].
\end{equation}

In the case where only the lowest Landau level is taken into account,
equation (78) reduces to the one that can be found in \cite{MTF2,MTF4},
namely, 
\begin{equation}
\mathbf{\nabla}^{2}\phi=4\frac{BT^{1/2}}{2^{3/2}\pi}I_{-1/2}\left(\frac{\mu+e\phi}{T}\right).\label{MG}
\end{equation}

Using the relation $\frac{d}{dx}I_{k}(x)=kI_{k-1}(x)$ we can obtain
a low $T$ expression for (\ref{MG}). Indeed, for low $T$ we can
write
\begin{eqnarray}
I_{-1/2}(x) & = & 2\frac{d}{dx}I_{1/2}(x)\approx2\frac{d}{dx}\left[\frac{2}{3}x^{3/2}\left\{ 1+\frac{3}{8x^{2}}\right\} \right]\nonumber \\
 & = & 2x^{1/2}\left\{ 1+\frac{3}{8x^{2}}\right\} -\frac{3}{2x^{3/2}}.\label{I1/2}
\end{eqnarray}

Then, at low T, equation (\ref{MG}) can be expanded as
\begin{equation}
\mathbf{\nabla}^{2}\phi=4\frac{B}{2^{3/2}\pi}\left[2(\mu+e\phi)^{1/2}\left\{ 1+\frac{3T^{2}}{8(\mu+e\phi)^{2}}\right\} -\frac{3T^{2}}{2(\mu+e\phi)^{3/2}}\right].\label{Lw T mg}
\end{equation}

To calculate the low-T Euler-Heisenberg correction to equation (\ref{MG})
we have to consider the Lagrangian (\ref{Low L}), this time taking
into account a magnetic term of the form $\mathbf{B}=B\widehat{\mathbf{k}}$
in the electromagnetic invariants (\ref{InvariantG}) and (\ref{InvariantF}).
With the magnetic terms included, the Gauss's law now reads
\begin{equation}
\nabla\cdot(A(T)E^{2}\mathbf{E}+\mathcal{B}(T)\mathbf{E}-E_{z}B^{2}\widehat{\mathbf{k}})=4\pi\rho,\label{lowTE-magnetic}
\end{equation}
where
\begin{equation}
\mathcal{B}(T)=B(T)+4\pi(4k-6c)B^{2}.
\end{equation}

From the form of the Gauss's law (\ref{lowTE-magnetic}), the modified
Poisson's equation
\begin{equation}
\nabla^{2}\phi+\mathfrak{D}_{EH-B}^{2}\phi=4\frac{BT^{1/2}}{2^{3/2}\pi}I_{-1/2}\left(\frac{\mu+e\phi}{T}\right),
\end{equation}
with
\begin{eqnarray}
\mathfrak{D}_{EH-B}^{2}\bullet & = &A(T)(\nabla\bullet)^{2}\nabla^{2}\bullet+2A(T)\left(\nabla\bullet\right)\cdot\left[\left(\nabla\bullet\right)\cdot\nabla\right]\nabla\bullet \nonumber\\
 &  & +4\pi(b+4kB^{2}-6cB^{2})\nabla^{2}\bullet+kB^{2}\frac{d^{2}\bullet}{dz^{2}}.
\end{eqnarray}

We can see that, due to existence of the magnetic field, the operator
$\mathfrak{D}_{EH-B}^{2}$ is not spherically symmetric.

In the procedure above there is a subtlety that we have to mention.
When substituting into the electromagnetic invariants we have considered
the magnetic field to be of the form $\mathbf{B}=B\widehat{\mathbf{k}}$.
However, an external magnetic field can induce the electric charges
to produce a magnetic field of their own \cite{inducedB,inducedB2}. So, in reality, the Gauss's
law has to take into account this induced field as well. However,
we have ignored the induced field since it will be much smaller than
the original external one.

\section{Applications}
In this section we remind the reader of some applications.
We will explicitly examine the details of an electric potential
in a neutral electron-positron plasma under conditions encountered in the
beginning of the universe. Secondly, we will recall how screening of
charges affects the alpha decay. 

\subsection{Ultra relativistic degenerate electron-positron gas}

The electron-positron plasma at an early stage of the Big-Bang presents
a situation where the thermal Euler-Heisenberg Lagrangian might prove
of great relevance. It is believed that the early pre-stellar period
of the evolution of the Universe was dominated by electrons and positrons 
having ultra relativistic temperatures \cite{misner}. In the time
between $10^{-6}s$ and $10s$ after the big bang, the universe reached
temperatures between $10^{9}K$ and $10^{13}K$ and was composed mainly
of electrons, positrons, and photons in thermodynamic equilibrium.
Furthermore, statistical mechanics states that for an electron-positron
plasma in an electrostatic field which is in equilibrium, the chemical
potential of the positrons and electrons must be the same in magnitude
at all points \cite{landau,zeldovich}.

Furthermore, in thermodynamic equilibrium the mean particle numbers
will change via the creation and annihilation processes, therefore
the total density $\eta_{-}-\eta_{+}$ will remain a constant. The
total charge density was calculated in \cite{electron-positron} and
can be written as
\begin{eqnarray}
e\eta_{-}-e\eta_{+} & = & \frac{\left(e\phi+\mu\right)}{3\hbar^{3}}\left[T^{2}+\frac{\left(e\phi+\mu\right)^{2}}{\pi^{2}}\right].\label{e-p charge density}
\end{eqnarray}

With the charge density (\ref{e-p charge density}) and the Euler-Heisenberg
contribution we can write for the potential the following equation
\begin{equation}
\left[\frac{8\alpha}{3}\ln\left(\frac{T}{m_{e}}\right)+\frac{4\alpha}{3}+1\right]\mathbf{\nabla}^{2}\phi=4\pi e\frac{\left(e\phi+\mu\right)T^{2}}{3\hbar^{3}}\left[1+\frac{\left(e\phi+\mu\right)^{2}}{\pi^{2}T^{2}}\right].\label{e+e-}
\end{equation}

With the change of variable $\Phi=\frac{e\phi+\mu}{T}$, the equation
(\ref{e+e-}) can be written as 
\begin{equation}
\mathbf{\nabla}^{2}\Phi=\frac{\Phi}{r_{EH}^{2}}\left[1+\Phi^{2}\right],
\end{equation}
where 
\begin{equation}
r_{EH}^{2}=\frac{(3/4\pi)(\hbar^{3}/e^{2}T^{2})}{\left[\frac{8\alpha}{3}\ln\left(\frac{T}{m_{e}}\right)+\frac{4\alpha}{3}+1\right]}.
\end{equation}

\subsection{Effect on tunneling probability}
The original Thomas-Fermi equation was derived for 
bound electrons. The derivation presented in this work 
shows that it is equally valid if the screening
happens in a gas of free electrons. We will use 
the Thomas-Fermi equation (\ref{TF}) in its simplest form, i.e.,
without the term proportional to $T^2$ and without Euler-Heisenberg 
corrections. It is evident that in equation (\ref{TF}) the 
length scale is given by the atomic Bohr radius whereas 
the important quantities entering the tunneling probability
of an alpha particle have to do with the much smaller nuclear scale.
Following \cite{Erma, Tiwary} one can expand the solution $\psi$
of the Thomas-Fermi equation which simplifies the calculations.  
According to (\ref{CV1}) we can write the interaction potential
between two positive charges (characterized by $Z_1$ and $Z_2$) as  
\begin{equation} \label{potential}
V(r)=\frac{Z_1Z_2 \alpha}{r}\psi(s)\,
\end{equation}
where $r = C s$ as used before. 
We will look for solutions of $\psi$ which at the lowest
order behave linearly, i.e., 
\begin{equation} \label{linear}
\psi_i(s) \simeq 1-d_i s\,.
\end{equation}
One such solution with a linear behaviour at the origin, which is one of the first attempts to
derive a semi-analytical solution of the Thomas Fermi equation, is given by
\cite{Baker} with $d_0$=1.588558. Other semi-analytical solutions 
\cite{Roberts, Csavinsky, Kesarwani, Oulne, Bougoffa} 
have been attempted and we list below
some of them in the order in which they are cited:
\begin{eqnarray} \label{solutions}
\psi_1(s)&=&(1+\eta\sqrt{s})e^{-\eta\sqrt{s}}\simeq =1-d_1s, \,\, d_1=3.6229 \nonumber \\
\psi_2(s)&=& (a_0e^{-\alpha_0 x} +b_0e^{-\beta_0 x})^2 \simeq 1-d_2x, \,\, d_2=1.2357 \nonumber \\
\psi_3(s)&=&(ae^{-\alpha x} +be^{-\beta x} + ce^{-\gamma x})^2 \simeq 1-d_3s, \,\, d_3=1.4042 \nonumber \\
\psi_4(s)&=&(1+A\sqrt{x} +Bxe^{-D\sqrt{x}})^2e^{-2A\sqrt{x}}\simeq 1-d_4s, \,\, d_4=1.45612 \nonumber \\
\psi_5(s)&=&\frac{1}{(1+A_0x)^2}\simeq 1-d_5s, \, \, d_5=0.9615
\end{eqnarray}
The potential for the alpha tunneling is in the first approximation given by a potential well modeling the
nuclear interaction plus the Coulomb or the modified Coulomb potential given in 
(\ref{potential}). In the semiclassical JWKB
approximation, the tunneling probability is simply given by \cite{Erma}, 
\begin{eqnarray} \label{prob}
P &\propto& e^{-\gamma} \nonumber \\
\gamma(E, r_1)&=&2\sqrt{2m}I(E, r_1)=2\sqrt{2m}\int_{r_1}^{r_2}\sqrt{[V(r)-E]}dr
\end{eqnarray}
where $m$ is the reduced mass, 
$r_1$ the first turning point given in our simple model by the
radius of the nucleus and
$r_2$ the second turning point determined by $V(r_2)=E$, with $E$ being the 
energy of the tunneling particle. 
In passing we note that we have omitted
some other approximate solutions which exist in the 
literature \cite{Desaix, Wu, Esposito}.

The integral $I$
with the Coulomb potential can be solved analytically to be \cite{Erma}, 
\begin{equation} \label{I}
I(E,r_1)=2\sqrt{2m}\frac{Z_1Z_2\alpha}{\sqrt{E}}\left[\cos^{-1}(x^{1/2}) -x^{1/2}(1-x)^{1/2}\right]
\end{equation}
with $x=(Er_1)/(Z_1Z_2\alpha)$.
Since the modification of the electromagnetic interaction brought by the Thomas-Fermi equation
can be approximated by $1-d_i s$, the correction to the potential is simply a constant.
The integral for the modified Coulomb problem is then $I(E^*, r_1)$ with 
$E^*=E+Z_1Z_2d_i\alpha/C$.
Correspondingly, we have $\gamma^*=\gamma(E^*, r_1)$. We have chosen the 
few examples (with exerimental $Q$-values \cite{nndc} denoted above as $E$)
with some of them being the same as in \cite{Erma}. 
The nuclear radii are taken from \cite{datatable}. 
In table 1 we summarize the effects in the form of the ratio 
of half-lives, $\tau/\tau^*$ for the decays, 
$^{106}_{52}$Te $\to \, ^4$He + $^{102}_{50}$Sn, 
$^{148}_{62}$Sm 
$\to\, ^{4}_{2}$He +  $^{144}_{60}$Nd, 
$^{222}_{86}$Rn $\to \,^{4}_2$He + $^{218}_{84}$Po, 
and 
$^{240}_{96}$Cm $\to \, ^4_2$He 
+ $^{236}_{94}$Pu. Though the exact values of half-lives 
(and hence also the screening effects) are 
sensitive to the Q-values \cite{Mandme}, the   
increase in the half-life due to screening seems to be quite sizable 
in some of the cases considered. The results prompt us to consider 
a more sophisticated calculation, with the following points in future:   
(i) Inclusion of the $T^2$ term
in the Thomas-Fermi equation for different gas temperatures, 
(ii) including the Euler-Heisenberg corrections and
(iii) improving the nuclear model such that the first turning point is 
also sensitive to the nuclear potential.
In passing we note that the Gamow factor 
$e^{-\gamma}$ appears also in stellar reaction rates $R$ defined by
\begin{equation} \label{R}
R \propto \int_0^{\infty}\, e^{-\gamma}S(E) e^{-E/kT}dE
\end{equation}
where $S(E)$ is the astrophysical S-factor \cite{Illiadis} which is sometimes 
approximated by a constant. It would be interesting to study the 
screening effects in the 
reaction rates (which eventually affect the abundance of elements) in the fusion 
reactions in stars within a more refined model as mentioned above. 
%It is then straightforward to calculate the ratio $R^*/R$ where $R^*$ refers to the modified Coulomb potential.
%Using the same approximations as above one obtains 
%$R^*_i/R_i =e^{D_i/kT}$ with $D=-Z_1Z_2\alpha d_i/C$.
%The relative error due to the different solutions can be estimated as $R^*_i/R^*_j=e^{\Sigma_{ij}}$
%with $\Sigma_{ij}\equiv Z_1Z_2\alpha(d_j-d_i)/(CkT)$. For $^{148}_{62}Sm$ and $d_i-d_j \sim 0.2$ we obtain
%$\Sigma_{ij} \sim 3\times 10^{-3} (MeV/T)$. 
%In other words, if $3\times 10^{-3} > T$ the ratio between the two
%approximate solutions will start growing exponentially. 
%Here again we see the necessity to refine the calculation. 
\begin{table}[ht]
\caption{The effect of electron gas on alpha tunneling. 
$\tau^*$ is the half-life of the decaying nucleus within the electron medium.}
\begin{tabular}{|l|l|l|l|l|}
  \hline
 $d_i$ & $\tau/\tau^*$ & $\tau/\tau^*$ & $\tau/\tau^*$ & $\tau/\tau^*$ \\
 & $^{106}_{52}$Te &$^{148}_{62}$Sm & $^{222}_{86}$Rn& $^{240}_{96}$Cm \\
   \hline
&&&&\\
0.962 & 1.123 &1.776 & 1.302 & 1.324 \\
1.236 & 1.161 &2.090 & 1.404 & 1.434 \\
1.404 & 1.185 &2.309 & 1.471 & 1.506 \\
1.456 & 1.192 &2.381 & 1.492 & 1.529 \\
1.589 & 1.211 &2.575 & 1.547 & 1.589 \\
3.630 & 1.547 &8.476 & 2.693 & 2.859 \\
\hline
\end{tabular}
\end{table}
 
\section{Conclusions}
The effect of surrounding matter at finite temperature on the electric potential
of an object is encoded in the Feynman-Metropolis-Teller equation (\ref{FMT}).
From this equation, 
various equations can be derived imposing different conditions on the matter.
Among the well known equations which emerge are the 
Thomas-Fermi and Poisson-Boltzmann equations.
Other, new equations like the relativistic Thomas-Fermi equation have 
been derived in the present work.
We have stressed the importance and the universal applicability of these equations.
Therefore, it appears timely to consider quantum corrections to these equations.
We have calculated these corrections using the Euler-Heisenberg theory at finite temperature.
For non-degenerate matter and high temperature analytical solutions have 
been presented.
Although our emphasis was on the derivations of these equations we have touched upon
two examples where it can be applied. One example concerns the electron-positron neutral plasma
under the Big-Bang conditions in the early universe. The other was a reminder of the state of art
of screening charges in astrophysics and its effect on alpha tunneling. The size of the effect
makes us think that a more detailed investigation including temperature effects and the
quantum corrections is in order. This will be attempted in a future publication.
As we already mentioned the applicability of the equations resulting from
the Feynman-Metropolis-Teller is manifold and not limited to the examples we presented here. 
Apart from atomic physics \cite{ElliotLieb}, plasma physics
\cite{Andres2} and biological applications \cite{Andres3}, one can also find
Thomas-Fermi like equations in gravitational physics \cite{Andres4}. 
Future projects could
probe into such equations replacing the Fermi-Dirac distribution by the corresponding Bose-Einstein for bosons.
Regarding the novel aspects where Thomas-Fermi equations could be used we mention graphene
where the electrons are treated relativistically \cite{graphene}.

With the inclusion of the quantum corrections we obtain a complete picture of 
the electric fields at finite temperature from which the electromagnetic force
can be easily calculated. Forces at finite temperature, of a different 
nature than the electromagnetic one, 
can, in general, be treated within quantum field theory at finite temperature (see \cite{we} for an example).

\section*{Appendix: Expansions for the charge density}

We review the form of the particle density for the limiting cases
of both non-relativistic and ultra relativistic particles. The special
case of ultra relativistic electron-positron plasma is shown at the
end. 

The quantity of interest is

\begin{eqnarray}
\eta(\mathbf{r},T) & = & 2\frac{(2\pi)^{3}}{\hbar^{3}}\int_{0}^{\infty}\frac{d\mathbf{p}}{e^{\beta\left(K+q\phi+\mu\right)}+1},\label{FD}
\end{eqnarray}
where $K$ is the kinetic energy of the particles.

For high temperatures the $+1$ in the denominator of (\ref{FD})
can be ignored. Under this consideration of non-degeneracy, the equation
(\ref{FD}) simplifies to
\begin{eqnarray}
\eta(\mathbf{r},T) & \approx & 2\frac{(2\pi)^{3}}{\hbar^{3}}e^{-\beta\left(q\phi+\mu\right)}\int_{0}^{\infty}p^{2}e^{-\beta K}dp.\label{MB}
\end{eqnarray}

Equation (\ref{FD}) can be simplified further by taking into account
the normalization condition

\begin{eqnarray}
N & = & \int\eta(\mathbf{r},T)d\mathbf{V}=2\frac{(2\pi)^{3}}{\hbar^{3}}e^{-\beta\mu}\int e^{-\beta(K+q\phi)}d\mathbf{p}d\mathbf{v},
\end{eqnarray}
where $N$ is the total number of particles. From the above we can
write for the chemical potential

\begin{equation}
e^{-\beta\mu}=\frac{N}{2\frac{(2\pi)^{3}}{\hbar^{3}}\int e^{-\beta(K+q\phi)}d\mathbf{p}d\mathbf{v}}.\label{chemP}
\end{equation}

Replacing (\ref{chemP}) into (\ref{MB}) we get

\begin{equation}
\eta(\mathbf{r},T)=\frac{Ne^{-q\phi\beta}\int e^{-\beta K}d\mathbf{p}}{\int e^{-\beta(K+q\phi)}d\mathbf{p}d\mathbf{v}}\approx\left(\frac{N}{V}\right)e^{-q\phi\beta}.\label{classical limit}
\end{equation}
In the last step of (\ref{classical limit}) we have made the final
approximation $\int e^{-q\phi\beta}d\mathbf{v}\approx V$, the total
volume. The justification is based on the assumption that for high
$T$ the exponential $e^{-q\phi\beta}$ will be small for almost all
the volume considered.

The approximation for the degenerate case involves a Sommerfeld's 
expansion in power series of $\frac{\mu+q\phi}{T}$ for the equation

\begin{equation}
\eta(\mathbf{r},T)=\frac{4(2\pi)^{4}}{\hbar^{3}}\int_{0}^{\infty}\frac{p^{2}dp}{e^{\beta\left(K+q\phi-\mu\right)}+1}.
\end{equation}

The first two terms for the non-relativistic cases read \cite{Feynman,Principles of stellar structure }

\begin{equation}
\eta(\mathbf{r},T)\approx\frac{2(2\pi)^{4}(2mT)^{3/2}}{\hbar^{3}}\left(\frac{2}{3}\left(\frac{q\phi-\mu}{T}\right)^{3/2}+\frac{\pi^{2}}{12}\left(\frac{q\phi-\mu}{T}\right)^{-1/2}\right).\label{non-rel degenerate}
\end{equation}

The ultra relativistic expansion is given by \cite{Principles of stellar structure }

\begin{equation}
\eta(\mathbf{r},T)\approx\frac{4(2\pi)^{4}}{\hbar^{3}}\left(q\phi-\mu\right)^{3}\left(1+\frac{3}{4}\pi^{2}\left(\frac{T}{q\phi-\mu}\right)^{2}\right).\label{ult-rel degenerate}
\end{equation}
A special case is the electron-positron plasma \cite{electron-positron}.
Due to the relation between their chemical potentials, the exact total
charge density $e\eta_{-}-e\eta_{+}$ can be written without any simplifying
assumption as

\begin{eqnarray}
e\eta_{-}-e\eta_{+} & = & \frac{4(2\pi)^{4}}{\hbar^{3}}\int_{0}^{\infty}dp\, p^{2}\left[\frac{1}{e^{\beta\left(p-e\phi-\mu\right)}+1}-\frac{1}{e^{\beta\left(p+e\phi+\mu\right)}+1}\right]\\
 & = & (2\pi)^{3}\frac{\left(e\phi+\mu\right)}{3\hbar^{3}}\left[T^{2}+\frac{\left(e\phi+\mu\right)^{2}}{\pi^{2}}\right].
\end{eqnarray}


\begin{thebibliography}{10}
\bibitem{TFatomic}
See S. Fl\"ugge, {\it Practical Quantum Mechanics}, Springer, Berlin 1971; J. C. Slater, {\it Quantum Theory of Atoms}, McGraw-Hill, New York 1960;
L. Spruch, Rev. Mod. Phys. {\bf 63} (1991) 151.
\bibitem{TFastro}
S. L. Shapiro and S. A. Teukolsky, {\it Black Holes, White Dwarfs and Neutron Stars}, 
Wiley, New York 1983.  
\bibitem{TFsolids}
J. C. Slater Rev. Mod. Phys. {\bf 6} (1934) 209; J. C. Slater and H. M. Krutter, 
Phys. Rev. {\bf 47} (1935) 559; A. Meyer and W. H. Young, J. Phys. C: Metal Phys, Suppl. 
{\bf 3} (1970) S348. 
\bibitem{PBplasma}
R. Ying and G. Kalman, Phys. Rev. {\bf A40}, 3927 (1989); 
M. Akbari-Moghanjoughi, Physics of Plasma {\bf 21}, 102702 (2014). 
\bibitem{PBsolutions}
M. Z. Bazant, M. S. Kilic and Ajdari, 
Advances in Colloid and Interface Science, {\bf 152}, 458 (2009). 
\bibitem{Feynman}R. P. Feynman, N. Metropolis, and E. Teller, Phys.
Rev. 75 , 1561 (1949).
\bibitem{EH}W. Heisenberg and H. Euler, Z. Phys. $\mathbf{98}$ (1936),
714.
\bibitem{weisskopf}V. Weisskopf, Mat.-Fis. Med. Dan. Vidensk. Selsk.
$\mathbf{14}$ (1936), 6.
\bibitem{Schwinger}J. Schwinger, Phys. Rev. $\mathbf{82}$ (1951),
664. 
\bibitem{A1}S. Adler, Ann. Phys. N.Y. $\mathbf{67}$ (1971), 599.
\bibitem{A2}Z. Bialynicka-Birula and I. Bialynicki-Birula, Phys.
Rev. D $\mathbf{2}$ (1970), 2341. 
\bibitem{Kruglov}S. I. Kruglov, Phys. Rev. D $\mathbf{75}$, (2007), 117301 
\bibitem{Costa} C. V. Costa, D. M. Gitman and A. E. Shabad, Phys. Scripta $\mathbf{90}$, 074012
(2015) 
\bibitem{Kruglov2}S. I. Kruglov, Mod. Phys. Lett. A, $\mathbf{32}$, 1750092 (2017)
\bibitem{A3}J.J. Klein and B.P. Nigam, Phys. Rev. $\mathbf{136}$
(1964), B1540.
\bibitem{A32}L. Labun and  J. Rafelski, Acta Phys. Pol. B 43, 2237 (2012). 
\bibitem{A4}R Battesti and C Rizzo 2013 Rep. Prog. Phys. $\mathbf{76}$
016401. 
\bibitem{A5}M. Marklund and J. Lundin, Eur. Phys. J. D {\bf 55} (2009) 
319. 
\bibitem{Dittrich -1}W. Dittrich and H. Gies, ``Probing the Quantum
Vacuum: Perturbative Effective Action Approach in Quantum Electrodynamics
and Its Applications'', Springer-Verlag-NY (2000). 
\bibitem{prapaper} A. D. Bermudez Manjarres and M. Nowakowski, 
Phys. Rev. A (2017) (in press). 
\bibitem{Dittrick first FT}W. Dittrich, Phys. Rev. D $\mathbf{19}$,
2385 (1979). 
\bibitem{FT2}P.H. Cox, W.S. Hellman and A. Yildiz, Ann. Phys. 154,
211 (1984). 
\bibitem{FT3}M. Loewe and J.C. Rojas, Phys. Rev. D $\mathbf{46}$,
2689 (1992). 
\bibitem{FT4}A.K. Ganguly, P.K. Kaw and J.C. Parikh, Phys. Rev. C
$\mathbf{51}$, 2091 (1995). 
\bibitem{strong field FT}I.A. Shovkovy, Phys. Lett. B $\mathbf{441}$,
313 (1998). 
\bibitem{FT5}A. Chodos, K. Everding and D.A. Owen, Phys. Rev. D 42,
2881 (1990). 
\bibitem{two loop}H. Gies, Phys. Rev. D 61, 085021 (2000). 
\bibitem{Hight T L}F.T. Brandt, J. Frenkel and J.C. Taylor, Phys.
Rev. D 50, 4110 (1994).
\bibitem{Hight T L2}F.T. Brandt and J. Frenkel, Phys. Rev. Lett.
74, 1705 (1995). 
\bibitem{PS }P. Elmfors and B.-S. Skagerstam, Phys. Lett. B 427,
197 (1998).
\bibitem{PS 2}V .Ch. Zhukovsky, T.L. Shoniya and P .A. Eminov, Zh.
Eksp. Teor. Fiz. 107, 299 (1995); J. Exp. Theor. Phys. 80, 158 (1995).
\bibitem{pair1}P. H. Cox, W .S. Hellman and A. Yildiz, Ann. Phys.
{\bf 154}, 211 (1984). 
\bibitem{pair2}J. Hallin and P. Liljenberg, Phys. Rev. D {\bf 46}, 2689
(1992). 
\bibitem{pair3}A.K . Ganguly, P .K . Kaw and J.C. Parikh, Phys.
Rev. C 51, 2091 (1995). 
\bibitem{Light 1}W. Dittrich and H. Gies, Phys. Rev. D 58, 025004
(1998). 
\bibitem{light 2}J.L. Latorre, P. Pascual and R. Tarrach, Nucl. Phys. B 437, 60 (1995).
\bibitem{light 3}H. Gies, Phys. Rev. D $\mathbf{60}$, 105033 (1999). 
\bibitem{electron-positron}N. L. Tsintsadze, A. Rasheed, H. A. Shah,
and G. Murtaza, Phys. Plasmas $\mathbf{16}$, 112307 (2009). 
\bibitem{Landau}B.V.Berestetskii, E.M.Lifshitz, and L.P.Pitaevskii,
\textquotedbl{}Quantum Electrodynamics\textquotedbl{}, Butterworth
- Heinemann, Oxford, 1999. 
\bibitem{Hansen}
C. J. Hansen, S. D. Kawaler {\it Stellar interiors-Physical Principles, Structure and
Evolution}, Springer, New York 1994. 
\bibitem{Bethe}R.E. Marshak, H. Bethe, Astrophys. J. 91 (1940) 239.
\bibitem{plasma DH}A. Piel, ''Plasma Physic:sAn Introduction to
Laboratory, Space, and Fusion Plasmas'', Springer-Verlag, Berlin
Heidelberg (2010), p. 35. 
\bibitem{Tsallis 1}C. Tsallis, J. Stat. Phys. 52 (1988) 479. 
\bibitem{tsallis euler turbulence}B.M. Boghosian, Phys. Rev. E 53
(1996) 4754. 
\bibitem{tsallis gravitation}A.R. Plastino, A. Plastino, Phys. Lett.
A 174 (1993) 384. 
\bibitem{tsallis ferro}P. Jund, S.G. Kim, C. Tsallis, Phys. Rev.
B 52 (1995) 50. 
\bibitem{tsallis n stars} D. Menezes, A. Deppman, E. Meg\'{i}as, L. Castro,
Eur. Phys. J. A 51 (2015) 155. 
\bibitem{bertulani}
S.Q. Hou, J.J. He, A. Parikh, D. Kahl, C.A. Bertulani, T. Kajino, G.J. Mathews, G. Zhao, Astrophys. J. {\bf 834} (2017) 165. 
\bibitem{Tsallis TM-1}E. Martinenko, B. K. Shivamoggi Phys. Rev.
A $69$, 052504 (2004). 
\bibitem{Tsallis RTF-1}K. Ourabah, M. Tribeche, Physica A 393 (2014)
470-474. 
\bibitem{Tsallis Numerics}D.F. Torres, U. TirnakliPhysica A 261 (1998)
499. 
\bibitem{MTF 1}B.B. Kadomtsev, Soviet Phys. JETP, $\mathbf{31}$
(1970), p. 945. 
\bibitem{MTF2}D.H. Constantinescu and G. Moruzzi, Phys. Rev. D 18
(1978) 1820. 
\bibitem{MTF3}S.H. Hill, P.J. Grout and N.H. March, J. Phys. B18
(1985) 4665. 
\bibitem{MTF4}B.K. Shivamoggi, P.P.J.M. Schram, Physica A 215 (1995)
387-39. 
\bibitem{MTF5}A. Thorolfsson, \"{O}. R\"{o}gnvaldsson, J. Yngvason. Astrophys.
J. $\mathbf{502}$ 847 (1998). 
\bibitem{inducedB}D.M Gitman and A.E. Shabad, Phys.Rev.D $\mathbf{86}$,
125028 (2012). 
\bibitem{inducedB2}T.C. Adorno, D.M. Gitman, A.E. Shabad, Eur. Phys.
J. C $74$, 2838 (2014); Phys. Rev. D $\mathbf{89}$, 047504 (2014). 
\bibitem{misner}W. Misner, K. S. Thorne, and J. A. Wheeler, ``Gravitation'',
Freeman, San Francisco, 1980, p. 764. 
\bibitem{landau}L. D. Landau and E. M. Lifshitz, ``Statistical Physics'',
Butterworth- Heinenann, Oxford, 1998, p. 315. 
\bibitem{zeldovich}Ya. B. Zel\textquoteright{}dovich and Y. P. Raizer,
Physics of Shock Waves and High- Temperature Hydrodynamic Phenomena
Academic, New York (1966), p. 222.
\bibitem{Erma}
V. Erma, Phys. Rev. {\bf 105} (1957) 1784.
\bibitem{Tiwary}
A. Jain and V. K. Tewary, Zetschrift f\"ur Astrphysics, {\bf 54} (1962) 107. 
\bibitem{Baker} 
E. B. Baker, Quart. Appl. Math. {\bf 36} (1930) 630. 
\bibitem{Roberts}
E. Roberts, Phys. Rev. {\bf 170} (1968) 8.
\bibitem{Csavinsky}
P. Csavinsky, Phys. Rev. {\bf A8} (1973) 1688. 
\bibitem{Kesarwani}
R. N. Kesarwani and Y. P. Varshni, Phys. Rev. {\bf A23} (1981) 991. 
\bibitem{Oulne}
M. Oulne, Int. Rev. Phys. {\bf 6} (2010) 349, 
ibid Applied Mathematics and Computation {\bf 228} (2011) 303. 
\bibitem{Bougoffa}
L. Bougoffa and R. Rach, Rom. J. Phys. {\bf 60} (2015) 1032. 
\bibitem{tsallis book}C. Tsallis, ``Introduction to nonextensive
statistical mechanics: approaching a complex world'', Springer-Verlag,
New York (2009). 
\bibitem{Desaix}
M. Desaix, D. Anderson and M. Lisak, Eur. J. Phys. {\bf 24} (2004) 699. 
\bibitem{Wu}
M. Wu, Phys. Rev {\bf A26} (1982) 57. 
\bibitem{Esposito}
S. Esposito, Am J. Phys. {\bf 70} (2002) 852.  
\bibitem{nndc}
Q-values obtained from the database at http://www.nndc.bnl.gov/qcalc/. 
\bibitem{datatable}
I. Angeli and K. P. Marinova, Atomic Data and Nuclear Data Tables {\bf 99} (2013) 
69.  
\bibitem{Mandme}
N. G. Kelkar and M. Nowakowski, J. Phys. G {\bf 43} (2016) 105102. 
\bibitem{Illiadis}
C. Iliadis, {\it Nuclear Physics of Stars}, Wiley-VCH 2007. 
\bibitem{ElliotLieb}
E. H. Lieb and B. Simon, Advances in Mathematics, {\bf 23} (1977) 22. 
\bibitem{Andres2} D. Michta , F. Graziani , and M. Bonitz , Contrib. 
Plasma Phys. 55, 437 (2015). 
\bibitem{Andres3}Fogolari F, Brigo A, Molinari H, J. Mol. Recognit. (2002), 15 377-392. 
\bibitem{Andres4}Bilic N and Viollier R D 1999 
Eur. Phys. J.C {\bf 11} 173.  
\bibitem{graphene}
A. H. Castro Neto at al., Rev. Mod. Phys. {\bf 81} (2009) 109. 
\bibitem{we} F. Ferrer, J.A. Grifols and M. Nowakowski, Phys. Rev. {\bf D61} 
(2000) 057304. 
\bibitem{Principles of stellar structure }John P. Cox, ``Principles
of stellar structure Volume II: Applications to stars'', Routledge
(1968), p. 801.
%\bibitem{tsallis fermi-dirac}J. Rozynek, Physica A 440, 27 (2015).
\end{thebibliography}
\end{document}